\documentclass[aip,amsmath,amssymb,reprint]{revtex4-1}
\usepackage{hyperref}
\hypersetup{
   breaklinks=true,   
}

\usepackage{graphicx}
\usepackage{dcolumn}
\usepackage{bm}

\usepackage[utf8]{inputenc}
\usepackage[T1]{fontenc}
\usepackage{mathptmx}
\usepackage{etoolbox}
\usepackage{siunitx}

\makeatletter
\def\@email#1#2{%
 \endgroup
 \patchcmd{\titleblock@produce}
  {\frontmatter@RRAPformat}
  {\frontmatter@RRAPformat{\produce@RRAP{*#1\href{mailto:#2}{#2}}}\frontmatter@RRAPformat}
  {}{}
}%
\makeatother

\begin{document}

\preprint{AIP/123-QED}

\title{Next-generation high transmission neutron optical devices utilizing micro-machined structures}

\author{C. Kapahi} 
\affiliation{Institute for Quantum Computing, University of Waterloo,  Waterloo, ON, Canada, N2L3G1}
\affiliation{Department of Physics, University of Waterloo, Waterloo, ON, Canada, N2L3G1}
\email{c3kapahi@uwaterloo.ca}
\author{D. Sarenac}
\affiliation{Institute for Quantum Computing, University of Waterloo,  Waterloo, ON, Canada, N2L3G1}
\author{M. Bleuel}
\affiliation{National Institute of Standards and Technology, Gaithersburg, Maryland 20899, USA}
\affiliation{Department of Materials Science and Engineering, University of Maryland, College Park, MD  20742-2115}
\author{D. G. Cory}
\affiliation{Institute for Quantum Computing, University of Waterloo,  Waterloo, ON, Canada, N2L3G1}
\affiliation{Department of Chemistry, University of Waterloo, Waterloo, ON, Canada, N2L3G1}
\author{B. Heacock}
\affiliation{National Institute of Standards and Technology, Gaithersburg, Maryland 20899, USA}
\author{M. Henderson} 
\affiliation{Institute for Quantum Computing, University of Waterloo,  Waterloo, ON, Canada, N2L3G1}
\author{M. G. Huber}
\affiliation{National Institute of Standards and Technology, Gaithersburg, Maryland 20899, USA}
\author{I. Taminiau}
\affiliation{Institute for Quantum Computing, University of Waterloo,  Waterloo, ON, Canada, N2L3G1}
\author{D. A. Pushin}
\affiliation{Institute for Quantum Computing, University of Waterloo,  Waterloo, ON, Canada, N2L3G1}
\affiliation{Department of Physics, University of Waterloo, Waterloo, ON, Canada, N2L3G1}

\date{\today}

\begin{abstract}
Neutrons have emerged as a unique probe at the forefront of modern material science, unrivaled in their penetrating abilities. A major challenge stems from the fact that neutron optical devices are limited to refractive indices on the order of $n\approx 1 \pm 10^{-5}$. By exploiting advances in precision manufacturing, we have designed and constructed a micro-meter period triangular grating with a high aspect ratio of $14.3$. The manufacturing quality is demonstrated with white-light interferometric data and microscope imaging. Neutron scattering experiment results are presented, showing agreement to refraction modelling. The capabilities of neutron Fresnel lenses based on this design are contrasted to existing neutron focusing techniques and the path separation of a prism-based neutron interferometer is estimated.
\end{abstract}

\maketitle

\section{Introduction}

Neutrons possess the unique ability to deeply penetrate materials due to their zero charge making them unparalleled probes of bulk material. In particular, neutron optics has established itself as an indispensable tool for measuring material properties and probing internal magnetic and phononic interactions~\cite{Werner2001NeutronMechanics, Cronin2009OpticsMolecules}. In Small Angle Neutron Scattering (SANS) experiments, a compound refractive lens (CRL) is often used to focus an incident neutron beam. CRL devices use a series of lenses, in some cases up to 30 individual elements~\cite{Eskildsen1998CompoundNeutrons, Beguiristain2002ALens, Gary2013CompoundApplications}. These devices, while effective, require large amounts of material to construct, resulting in a significant amount of beam attenuation. Alternatives to neutron CRL devices which circumvent these limitations include lenses that utilize magnetic fields~\cite{Oku2000NeutronPrism, Yamada2015DevelopmentSextupole, Shimizu1999MeasurementMagnet, Shimizu1997ColdGradient, Shimizu2000MagneticLens, Shimizu2002DevelopmentDevices, Littrell2007MagneticBeams}, focusing capillaries~\cite{Arzumanov2016ACapillaries, Kumakhov1992ALens}, Wolter mirrors~\cite{Wu2017WolterImaging}, and Fresnel lenses~\cite{Oku2001DevelopmentOptics}.

In addition to SANS methods, neutron optics has expanded to include a wealth of techniques from polarized neutron experiments~\cite{Kardjilov2008Three-dimensionalNeutrons, Jericha2007ReconstructionTomography, Kageyama2009DevelopmentImaging}, single-crystal interferometers~\cite{Pushin2015NeutronTechnology, Sarenac2016HolographyInterferometer, Lemmel2015NeutronFields, Hasegawa2003ViolationInterferometry, Bartosik2009ExperimentalInterferometry}, and grating-based neutron interferometers~\cite{Sarenac2018ThreeApplications, Pushin2017Far-fieldImaging, Heacock2019AngularVisibility}. The first neutron interferometer was built with two prisms that caused Fresnel interference by means of wave front division~\cite{maier1962interferometer}. With this device a path separation of only $\SI{60}{\um}$ was obtained, which was later improved to a few centimeters with a perfect crystal Mach-Zehnder interferometer relying on Bragg diffraction \cite{Rauch1974TestInterferometer}. The perfect crystal interferometer allowed for a number of investigations on neutron properties and interactions with unprecedented precision~\cite{Rauch1975VerificationFermions, werner1988neutron, clark2015controlling, Li2016NeutronField, denkmayr2014observation} but suffered some important drawbacks including a narrow wavelength acceptance and difficulty of fabrication~\cite{arif1994multistage, saggu2016decoupling, shahi2016new, zawisky2010large}.

Utilizing a diamond turning machine, we have fabricated a high aspect ratio structure of refractive prisms that deflects a neutron beam. In this work, we demonstrate a functioning triangular array refractive prism (TARP) with neutrons at the BT-5 Ultra Small Angle Neutron Scattering (USANS) instrument located at the National Institute for Standards and Technology (NIST) Center for Neutron Research (NCNR) facility~\cite{Barker2005DesignNIST}. Experimental results are compared to theoretical predictions based on Snell's Law. The TARP functions as a Fresnel prism and enables prism-based neutron interferometers to create path separations larger than previously possible. The micro-prisms used in this TARP device enable the construction of next-generation neutron focusing devices.

\section{Micro-Prism Fabrication}

\begin{figure*}
    \centering
    \includegraphics[width=0.9\textwidth]{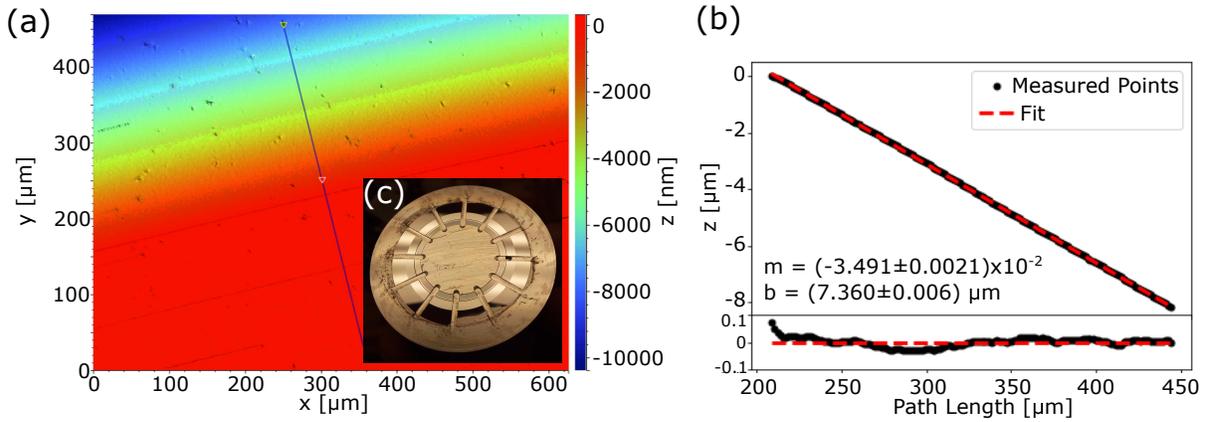}
    \caption{(a) White light interferometry data showing the machined surface of the prism segment before it was cut and assembled into the TARP. A false-colour representation is used to indicated surface height. Surface imperfections can be observed which correspond to a surface finish of 20 nm Ra. Points along the blue line were sampled to measure the angle of this surface. Note that the base of the prism is visible in this image, shown in red. (b) These points show an angle of $1.999^\circ \pm 0.001^\circ$. This angle matches the designed surface angle of $2^\circ$. The residual plot shown below indicates that within this region, surface form is straight to within $\SI{0.1}{\micro\meter}$. (c) Micro-prisms before assembly into the TARP. Here the angled faces of each prism can be seen tapering radially and each prism has been segmented in equal pieces. These pieces are then assembled with the non-angled section at the base of each prism.}
    \label{fig:white_interferometer_fig}
\end{figure*}

High aspect ratio prisms can now be realized thanks to an ultra high precision machining center (Moore Nanotech 350FG~\cite{NIST_disclaimer}) with C-axis control and live tooling. The choice of material, fully annealed 5052-O aluminum, was dictated by its strength and formability coupled with the material's resistance to heat, allowing the piece to be mounted and unmounted using a low temperature wax. The individual prisms were fabricated from a 5052-O aluminum ring of $\SI{38}{\milli\meter}$ outer diameter (OD), $\SI{25}{\milli\meter}$ inner diameter (ID), and thickness of $\SI{250}{\micro\meter}$ that is held on a specially designed aluminum chuck that accommodates mounting by low temperature wax, which can alternatively be fully dissolved in acetone (Crystal bond 509-3). The back side of the ring is then turned using diamond tooling to create a flat reference surface with a surface finish of $\SI{20}{\nano\meter}$ roughness average (Ra) and form of better than $\SI{1}{\micro\meter}$. The ring is then remounted with the reference face facing the custom chuck to ensure a seating of better than $\SI{3}{\micro\meter}$ parallelism with the chuck face, which can be diamond turned flat each time if needed. The remounted ring is then segmented into pieces $\approx\SI{5}{\milli\meter}$ wide at their base by an auxiliary milling spindle, the segmented ring is shown in Figure~\ref{fig:white_interferometer_fig}(c). The finishing operation involves diamond turning a flat face with a $\SI{140}{\micro\meter}$ target thickness from $\SI{25}{\milli\meter}$ ID to $\SI{29}{\milli\meter}$ OD transitioning into a conical face, starting at $\SI{28}{\milli\meter}$ OD and tapering to zero thickness at $\SI{37}{\milli\meter}$ OD.

The angle of the radial cross-section of each segment equals 2$^\circ$ $\pm$0.02$^\circ$ with an uncertainty from two sources. The first error encountered arises from the parallelism of the front and back face which was measured by the maximum variation in thickness ($\SI{3}{\micro\meter}$) at the OD of $\SI{38}{\milli\meter}$ after facing a ring without conical feature. This results in an uncertainty of 0.005$^\circ$. A larger error arises from the fact that each prism is a segment of a truncated cone and thus an approximation of a true geometrical prism. Since the base of each segment would be the maximum usable area, and both sides of the fan-shaped segment would be cropped perpendicular to the base, it would reveal a hyperbolic curve which can be fitted with a straight line with an angle of 1.98$^\circ$. Therefore, the maximum error due to the conical shape of each segment is determined to be 0.02$^\circ$ at the region furthest from the center line. The most unpredictable error in the deviation from a perfect prism is due to the extremely thin apex of the prisms, which deforms at the slightest pressure. Heating the chuck to release the segments showed a severe buckling of all apices likely attributed to uneven rate of thermal expansion. Therefore, the piece was unmounting by dissolving the mounting wax in acetone instead of raising the temperature. After releasing the segments and gently cleaning, without agitation, the segments were carefully assembled in a custom full aluminum jig, which allowed the stack of prisms to be squeezed together by a fine pitch screw, creating an array of prism segments.

\begin{figure*}
    \centering
    \includegraphics[width=0.9\textwidth]{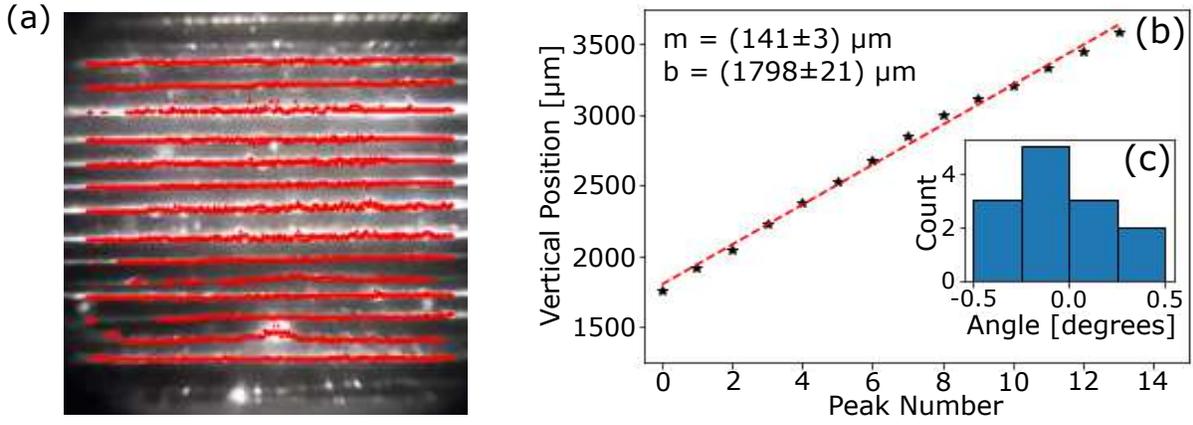}
    \caption{(a) Microscope image of assembled TARP device with peak locations labelled with red-dots. Peak locations along a single prism are fit to a line which is then used to determine prism spacing and alignment. (b) The y-intercepts of each linear fit from (a) are plotted and fit to a line. The slope of this line corresponds to the average spacing the prism peaks. (c) Histogram of the slope of fit lines from (a). This shows that prism alignment has a standard deviation of $\pm 0.3^\circ$.}
    \label{fig:microscope_images}
\end{figure*}

The manufacturing quality of the TARP was first evaluated using white-light interferometry and optical microscopy. The white-light interferometry data, shown in Figure~\ref{fig:white_interferometer_fig}(a), indicates the surface angle of annulus is $1.999^\circ \pm 0.001^\circ$. Additionally, the interferogram of the machined surface shows that the prism slope is straight and uniform. Once this annulus is cut and assembled, the alignment and spacing of prisms in the TARP can be measured by comparing the separation of the prism apices at different locations on the optical microscope image. Figure~\ref{fig:microscope_images}(a) shows red points where the prism peak is detected in software. Each series of red points corresponding to a single prism are then linearly fit to extract a series of slopes, $m_i$, and intercepts, $b_i$. The y-intercepts of these lines can be used to compare the spacing between each prism. Figure~\ref{fig:microscope_images}(b) shows these y-intercepts, $b_i$, which are again linearly fit to determine that the spacing between prism peaks is $\SI{141}{\micro\meter} \pm \SI{3}{\micro\meter}$. The slopes, $m_i$, show a variation in the alignment of these prisms with a standard deviation of $\pm 0.3^\circ$. A histogram of these slope values can be seen in Figure~\ref{fig:microscope_images}(c). The high aspect ratio, $14.3$, suggests that the TARP prototype would be capable of deviating a neutron beam as if it were a $\SI{3}{\milli\meter}$ wide prism that is $\SI{85.9}{\milli\meter}$ in height.

The refractive characteristics of the TARP were tested using cold neutrons at the BT-5 Ultra Small Angle Neutron Scattering (USANS) instrument located at the NCNR facility~\cite{Barker2005DesignNIST}. The experimental arrangement is shown in Figure~\ref{fig:exp}(a). 

\section{Device Modelling}

\begin{figure}
    \centering
    \includegraphics[width=0.45\textwidth]{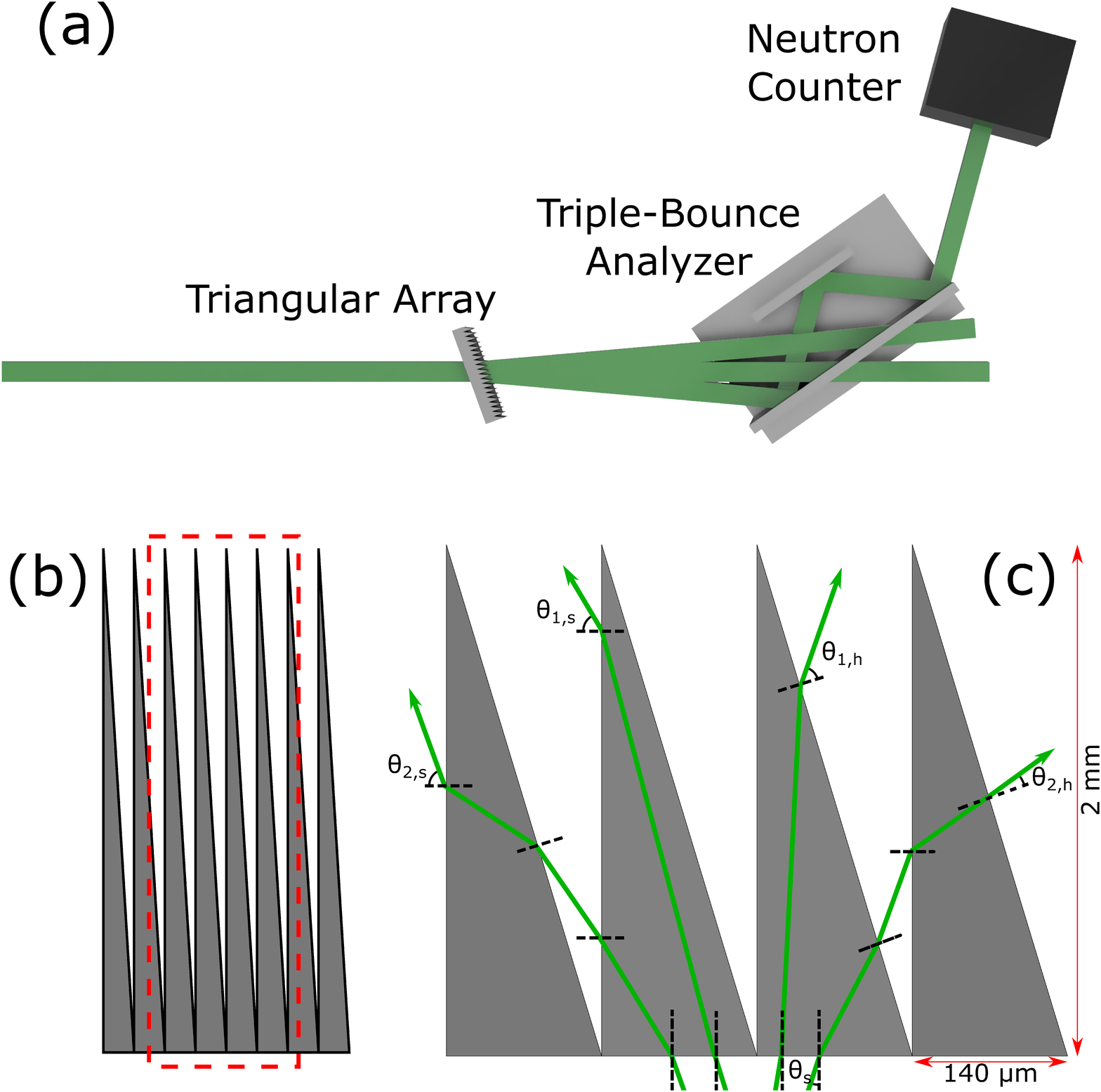}
    \caption{(a) Experimental apparatus used to measure beam deviation from the TARP. Small angle deviations are measured by utilizing a crystal applying Bragg diffraction to select a particular deviation angle caused by the sample. The neutron beam had a wavelength of $\SI{2.38}{\angstrom} \pm \SI{0.051}{\angstrom}$, accounting for the effect of slits in our experiment. (b) Scaled schematic of TARP, red dotted box indicated region which is magnified and shown with exaggerated scale in (c). Aspect ratio shown here ($> 14$) is correct for the prototype TARP. (c) Schematic of the TARP showing the arrangement and length scale of each prism. The neutron path and corresponding outgoing angle of the four possibilities modelled in Equation~\ref{eqn:approx} are indicated as green lines. Neutron path angles are exaggerated for illustrative purposes.}
    \label{fig:exp}
\end{figure}

The neutron ray deflection was calculated by propagating rays across material boundaries using Snell's Law. Taylor expanding these equations about $n=1$, we can obtain the well known approximation for prism deflection~\cite{wagh2004geometric}. Applying this to each side of the first and second prism that a neutron encounters, we find that the angles labeled in Figure~\ref{fig:exp}(c), where $\Delta\theta_{h,1}$ is the deviation of neutron ray that exits from the hypotenuse the first prism it encounters,

\begin{subequations}
\begin{align}
    \begin{split}
    \Delta\theta_{h,1} &\approx (1-n)[\tan(\pi/2-\beta-\theta) + \tan(\theta)],
    \end{split}\\
    \begin{split}
    \Delta\theta_{s,1} &\approx (1-n)[\tan(\pi/2-\theta) + \tan(\theta)],
    \end{split}\\
    \begin{split}
    \Delta\theta_{h,2} &\approx (1-n)[\tan(\beta-\pi/2-\theta-\Delta\theta_{h,1})\\
    &\hspace{1.5cm}+ \tan(\pi/2+\theta+\Delta\theta_{h,1})],
    \end{split}\\
    \begin{split}
    \Delta\theta_{s,2} &\approx (1-n)[\tan(-\pi/2+\theta+\Delta\theta_{s,1})\\
    &\hspace{1.5cm}+ \tan(\pi/2-\theta-\Delta\theta_{s,1}+\beta)],
    \end{split}
\end{align}
\label{eqn:approx}
\end{subequations}

\noindent
for rays incident at an angle, $\theta$, to a prism of angle, $\beta$.

Using these approximations, we can model the TARP with neutron rays in a Monte Carlo simulation. These rays have a wavelength and initial angle that is drawn from a Gaussian distribution, corresponding to the BT-5 instrument's wavelength range and angle distribution by the Darwin width of the monochromator, respectively~\cite{Barker2005DesignNIST}. Neutron rays possess an additional angle, correlated to its wavelength, labelled $\lambda_D$, caused by monochromator which functions via Bragg diffraction. From here the rays are propagated across each prism boundary, with Snell's Law dictating the angular deflection. Neutrons may also encounter adjacent prisms after exiting the first prism. The resulting beam deviation is then the difference between the final propagation angle of each ray and the incident ray angle. As the analyzing crystal also operates via Bragg diffraction, the angle $\theta_D$ is also subtracted from the final angle of each ray. This procedure is repeated at each sample angle with the Gaussian distribution of initial ray angles shifted by the sample angle.

\section{Experimental Results}

\begin{figure*}
    \centering
    \includegraphics[width=0.9\textwidth]{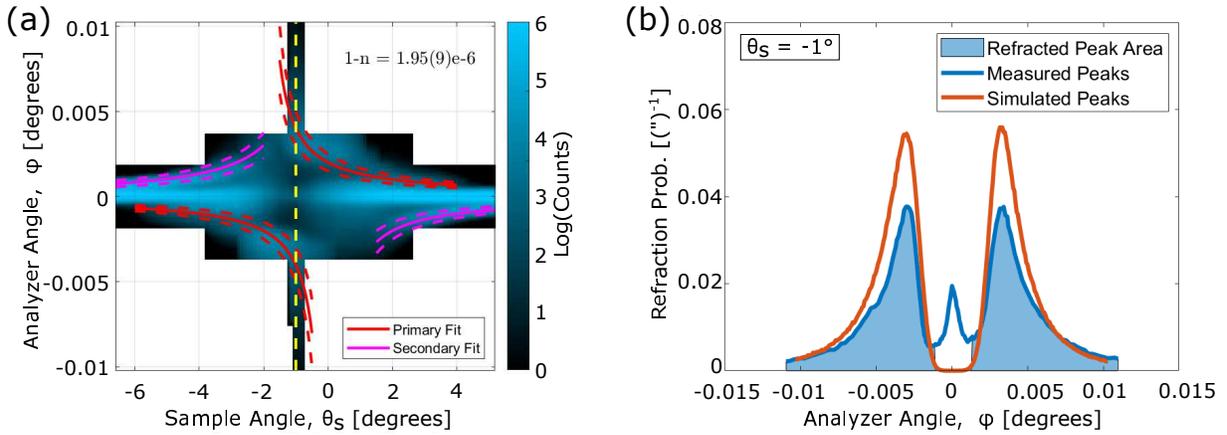}
    \caption{(a) USANS results from the TARP where false-colour corresponds to the logarithm of counts. Scattering peaks have been fit to a Gaussian to extract the peak location. Red and pink lines correspond to first and second order theoretical scattering curves, respectively, fit to these peaks. Curves are calculated via Equation~\ref{eqn:approx}, accounting for the BT-5 wavelength distribution, and fit with a free fit parameter for material index. Dotted lines show the $95\thinspace\%$ confidence interval of this fit. This fit returns a material index of $n = 1 - (1.95\pm0.09)\cdot 10^{-6}$. Dotted yellow line indicates a sample angle of $\theta_s = -1^\circ$. (b) Comparison of the TARP USANS measurements (blue) with Monte Carlo simulation (red) taken at $\theta_s = -1^\circ$. Shaded area in blue denotes region where neutron deflection is predicted by simulation. This area is used to determine the refraction efficiency of the TARP. Neutron ray deviations from USANS experiments with the TARP show qualitative agreement with simulations. Features of the experimental scattering peaks including the Lorentzian-like tails of the diffraction orders can be seen in simulation by considering misalignment between each prism within the TARP.}
    \label{fig:2018fits_sim}
\end{figure*}

In order to extract meaningful values from a fit of the approximate deviation in Equation~\ref{eqn:approx}, the sample angle axis of the experimental data must be aligned with $\theta_s = \ang{0}$ as defined by the device. This can be achieved by comparing experimental plots of constant sample angle, $\theta_s$, to simulation. At $\theta_s=\ang{-1}$, the primary scattering peaks are unique in the sense that they are of equal magnitude and can therefore be used to align the data. The comparison of simulation to experiment shown in Figure~\ref{fig:2018fits_sim}(b) allows us to specify this point in experiment as $\theta_s=\ang{-1} \pm \ang{0.12}$. Furthermore, the Lorentzian-like tails of the distribution can be recovered from simulation by integrating over a distribution of prism alignments. The effective wavelength distribution and divergence angle of the instrument, including apertures used in the experiment, can be measured by fitting the peak shape at $\theta_s=\ang{-1}$. This results in a wavelength range of $\SI{2.38}{\angstrom} \pm \SI{0.051}{\angstrom}$ and initial ray angle of $\pm (0.38\cdot10^{-3})^\circ$. The central peak seen in experiment is due to neutrons which were not deflected by the TARP.

Figure~\ref{fig:2018fits_sim}(a) shows a false-colour representation of the neutron intensity from the single-pixel detector at each sample angle, $\theta_s$, and Bragg analyzer angle, $\phi$. From the data, a central peak along $\phi=\ang{0}$ can be seen at each sample angle, with two roughly symmetric peaks diverging toward either positive or negative analyzer angles, $\phi$. Four peaks can be observed in total, with the two primary peaks nearest to $\theta_s=-1^\circ$ arising due to refraction from a single prism, while the secondary peaks arise from neutron rays that exit the initial prism and refract within the adjacent prism.

Agreement between theory and experiment can be demonstrated by comparing the theoretical material index of refraction to the value obtained by fitting Equation~\ref{eqn:approx} with a free parameter $\delta=1-n$. The data was fit as shown in Figure~\ref{fig:2018fits_sim}(a) and the resulting material index value is $n = 1 - (1.95\pm0.09)\cdot 10^{-6}$, which is in $95\thinspace\%$ confidence agreement with the expected value of $1 - n=1.93\cdot 10^{-6}$ for the aluminum alloy used in these prisms.

The simulated beam deviation is shown overlaying the experimental data in Figure~\ref{fig:2018fits_sim}(b). From this, we can see qualitative agreement between the simulated prisms and our experiment using the theoretical material index. These simulated peaks show that there exists a fraction of the beam whose deviation is not predicted by our model. We can define the relative efficiency of this manufactured TARP as the ratio between the peak area which overlaps with our simulation (shaded area in Fig.~\ref{fig:2018fits_sim}(b)) and the total area of the measured refraction peak. Therefore we find that at $\theta_s=-1^\circ$ our device has a relative efficiency of $\eta_{-1^\circ} = 86.3\thinspace\%$. Two other sample angles of interest are when $\theta_s = \beta$, $\eta_{-2^\circ} = 63.9\thinspace\%$, and when the sample is perpendicular to the beam, $\eta_{0^\circ} = 47.0\thinspace\%$. The primary source of these discrepancies is the central intensity, seen at $\phi = 0^\circ$ in Figure~\ref{fig:2018fits_sim}(b), suggesting that a significant fraction of the beam is not deflected by the TARP. There are many possible sources of this peak including neutrons which pass between prisms in the TARP and neutrons which exit the TARP through the flat area at the apex of each prism.

\section{Discussion}

We have designed and fabricated a neutron optical device which utilizes an array of high aspect ratio micro-prisms to deviate a neutron beam by large angles with high efficiency. The TARP has been modelled with Snell's Law and experimental data has been fit to measure material index of refraction, which agrees with the accepted value for aluminum 5052-O. The fabrication methods described here can be applied to a refractive prism neutron interferometer or to construct a Fresnel lens with micro-prisms. 

The efficiency of $\eta_{-2^\circ} = 63.9\thinspace\%$ is comparable to current state-of-the-art neutron optical devices, such as the thin film Fresnel lenses shown in~\cite{Oku2001DevelopmentOptics}, which achieved a focusing efficiency of $65\thinspace\%$. The CRL device in~\cite{Eskildsen1998CompoundNeutrons} achieved a transmission rate of $77\thinspace\%$, and a gain in intensity of 15. Another compound neutron lens that utilizes magnetic lenses achieved a gain of 37.5 but is restricted to polarized neutrons~\cite{Shimizu2000MagneticLens}. A TARP with prism angles between $1^{\circ}$ and $2.61^{\circ}$, could act as a neutron lens with focal length $f = \SI{2.77}{\meter}$ and spot size of $\SI{1}{\milli\meter}$, achieving an intensity gain of $>$100.

The beam deviation offered by the TARP can also be used to construct a neutron interferometer. Early experiments in neutron interferometry made use of refractive prisms to coherently separate $\SI{4.4}{\angstrom}$ neutrons by $\SI{60}{\micro\meter}$ over a $\SI{10}{\meter}$ flight path~\cite{maier1962interferometer}. Utilizing the TARP examined in this work, a separation of $\SI{870}{\micro\meter}$ is achievable with the $\SI{2.38}{\angstrom}$ neutrons used to test the sample, implying that a separation of $\SI{3}{\milli\meter}$ is realizable with $\SI{4.4}{\angstrom}$ neutrons over $\SI{10}{\meter}$. While the coherence properties of beams refracted by the TARP have not been studied, a neutron interferometer with a beam separation of $\SI{60}{\micro\meter}$ could be achieved with a beam-line of less than $\SI{1}{\meter}$ using two TARP devices similar to what was shown here.

\begin{acknowledgments}
This work was supported by the Canadian Excellence Research Chairs (CERC) program, the Natural Sciences and Engineering Research Council of Canada (NSERC) Discovery program, and the Collaborative Research and Training Experience (CREATE) program, the Canada First Research Excellence Fund (CFREF). Access to BT5 USANS Instrument was provided by the Center for High Resolution Neutron Scattering, a partnership between the National Institute of Standards and Technology and the National Science Foundation under Agreement No. DMR-1508249. We acknowledge the support of the National Institute of Standards and Technology, U.S. Department of Commerce, in providing the neutron research facilities used in this work and US Department of Energy, Office of Nuclear Physics, under Interagency Agreement 89243019SSC000025.
\end{acknowledgments}

\section*{Data Availability Statement}

The data that support the findings of this study are available from the corresponding author upon reasonable request.

\bibliography{ref}

\end{document}